\documentclass[aps,prd,groupedaddress,superscriptaddress, twocolumn]{revtex4-2}

\pdfoutput=1

\usepackage[separate-uncertainty,retain-explicit-plus,per-mode=symbol,binary-units]{siunitx}

\usepackage[T1]{fontenc}
\usepackage{hyperref}
\usepackage{xcolor}
\usepackage{graphicx}
\usepackage{array,booktabs}
\usepackage{float}
\usepackage{array,mathtools,dcolumn}
\usepackage{amsmath,amssymb}
\usepackage{multirow} 
\usepackage{tabularx}
\usepackage{booktabs}
\usepackage{comment}
\usepackage[normalem]{ulem}
\usepackage[caption=false]{subfig}
\usepackage[version=4]{mhchem}
\usepackage{xspace}
\usepackage[normalem]{ulem}
\usepackage{graphicx}
\usepackage[english]{babel}
\newcommand{\Qti}{Q_y^{\rm TI}}        

\usepackage{graphicx}
\usepackage{dcolumn}
\usepackage{bm}

\begin{document}



\title{\textbf{Sensitivity to low-mass WIMPs  with an improved liquid argon ionization response model within the DarkSide programme}}





\newcommand{\Alberta}{Department of Physics, University of Alberta, Edmonton, AB T6G 2R3, Canada}
\newcommand{\APC}{APC, Universit\'e de Paris Cit\'e, CNRS, Astroparticule et Cosmologie, Paris F-75013, France}
\newcommand{\AQLNGS}{INFN Laboratori Nazionali del Gran Sasso, Assergi (AQ) 67100, Italy}
\newcommand{\AQGSSI}{Gran Sasso Science Institute, L'Aquila 67100, Italy}
\newcommand{\AUM}{InstitutodeF\'õsica,Universidad Nacional Auto\'nomade M\'exico(UNAM), M\'exico 01000, Mexico}
\newcommand{\Augustana}{Physics Department, Augustana University, Sioux Falls, SD 57197, USA}
\newcommand{\Belgorod}{Radiation Physics Laboratory, Belgorod National Research University, Belgorod 308007, Russia}
\newcommand{\BHSU}{School of Natural Sciences, Black Hills State University, Spearfish, SD 57799, USA}
\newcommand{\BINP}{Budker Institute of Nuclear Physics, Novosibirsk 630090, Russia}
\newcommand{\Birmingham}{School of Physics and Astronomy, University of Birmingham, Edgbaston, B15 2TT, Birmingham, UK}
\newcommand{\BNL}{Brookhaven National Laboratory, Upton, NY 11973, USA}
\newcommand{\BOCentroFermi}{Museo Storico della Fisica e Centro Studi e Ricerche Enrico Fermi, Roma 00184, Italy}
\newcommand{\BOINFN}{INFN Bologna, Bologna 40126, Italy}
\newcommand{\BOUniPHY}{Physics Department, Universit\`a degli Studi di Bologna, Bologna 40126, Italy}
\newcommand{\CAUniCHE}{Department of Mechanical, Chemical, and Materials Engineering, Universit\`a degli Studi, Cagliari 09042, Italy}
\newcommand{\CAUniPHY}{Physics Department, Universit\`a degli Studi di Cagliari, Cagliari 09042, Italy}
\newcommand{\CAINFN}{INFN Cagliari, Cagliari 09042, Italy}
\newcommand{\Carleton}{Department of Physics, Carleton University, Ottawa, ON K1S 5B6, Canada}
\newcommand{\Campinas}{Physics Institute, Universidade Estadual de Campinas, Campinas 13083, Brazil}
\newcommand{\CentroFermi}{Museo della fisica e Centro studi e Ricerche Enrico Fermi, Roma 00184, Italy}
\newcommand{\Chicago}{Department of Physics and Kavli Institute for Cosmological Physics, University of Chicago, Chicago, IL 60637, USA}
\newcommand{\CIEMAT}{CIEMAT, Centro de Investigaciones Energ\'eticas, Medioambientales y Tecnol\'ogicas, Madrid 28040, Spain}
\newcommand{\Cluj}{National Institute for R\&D of Isotopic and Molecular Technologies, Cluj-Napoca, 400293, Romania}
\newcommand{\Columbia}{Physics Department, Columbia University, New York, NY 10027, USA}
\newcommand{\CTINFN}{INFN Catania, Catania 95121, Italy}
\newcommand{\CTUNI}{Universit\`a of Catania, Catania 95124, Italy}
\newcommand{\CTLNS}{INFN Laboratori Nazionali del Sud, Catania 95123, Italy}
\newcommand{\ENUniCEE}{Engineering and Architecture Faculty, Universit\`a di Enna Kore, Enna 94100, Italy}
\newcommand{\ETHZ}{Institute for Particle Physics, ETH Z\"urich, Z\"urich 8093, Switzerland}
\newcommand{\FNAL}{Fermi National Accelerator Laboratory, Batavia, IL 60510, USA}
\newcommand{\FortLewis}{Department of Physics and Engineering, Fort Lewis College, Durango, CO 81301, USA}
\newcommand{\GEUni}{Physics Department, Universit\`a degli Studi di Genova, Genova 16146, Italy}
\newcommand{\GEINFN}{INFN Genova, Genova 16146, Italy}
\newcommand{\GlenEllyn}{Glen Ellyn, Illinois 60137, USA}
\newcommand{\Hawaii}{Department of Physics and Astronomy, University of Hawai'i, Honolulu, HI 96822, USA}
\newcommand{\Houston}{Department of Physics, University of Houston, Houston, TX 77204, USA}
\newcommand{\IHEP}{Institute of High Energy Physics, Beijing 100049, China}
\newcommand{\INFN}{Istituto Nazionale di Fisica Nucleare, Roma 00186, Italia}
\newcommand{\IPNO}{Institut de Physique Nucl\`eaire dÕOrsay, 91406, Orsay, France}
\newcommand{\INSTM}{Interuniversity Consortium for Science and Technology of Materials, Firenze 50121, Italy}
\newcommand{\IPHC}{IPHC, Universit\'e de Strasbourg, CNRS/IN2P3, Strasbourg 67037, France}
\newcommand{\JINR}{Joint Institute for Nuclear Research, Dubna 141980, Russia}
\newcommand{\Krakow}{M. Smoluchowski Institute of Physics, Jagiellonian University, 30-348 Krakow, Poland}
\newcommand{\Kurchatov}{National Research Centre Kurchatov Institute, Moscow 123182, Russia}
\newcommand{\Lancaster}{Physics Department, Lancaster University, Lancaster LA1 4YB, UK}
\newcommand{\Laurentian}{Department of Physics and Astronomy, Laurentian University, Sudbury, ON P3E 2C6, Canada}
\newcommand{\Liverpool}{Department of Physics, University of Liverpool, The Oliver Lodge Laboratory, Liverpool L69 7ZE, UK}
\newcommand{\LNFINFN}{INFN Laboratori Nazionali di Frascati, Frascati 00044, Italy}
\newcommand{\LNLINFN}{INFN Laboratori Nazionali di Legnaro, Legnaro (Padova) 35020, Italy}
\newcommand{\Lodz}{Institute of Applied Radiation Chemistry, Lodz University of Technology, 93-590 Lodz, Poland}
\newcommand{\LPNHE}{LPNHE, CNRS/IN2P3, Sorbonne Universit\'e, Universit\'e Paris Diderot, Paris 75252, France}
\newcommand{\Manchester}{The University of Manchester, Manchester M13 9PL, United Kingdom}
\newcommand{\MendeleevUniverisity}{Mendeleev University of Chemical Technology, Moscow 125047, Russia}
\newcommand{\MEPhI}{National Research Nuclear University MEPhI, Moscow 115409, Russia}
\newcommand{\MIBIINFN}{INFN Milano Bicocca, Milano 20126, Italy}
\newcommand{\MIINFN}{INFN Milano, Milano 20133, Italy}
\newcommand{\MIPoliICA}{Civil and Environmental Engineering Department, Politecnico di Milano, Milano 20133, Italy}
\newcommand{\MIPoliCHE}{Chemistry, Materials and Chemical Engineering Department ``G.~Natta", Politecnico di Milano, Milano 20133, Italy}
\newcommand{\MIPoliEIB}{Electronics, Information, and Bioengineering Department, Politecnico di Milano, Milano 20133, Italy}
\newcommand{\MIPoliENE}{Energy Department, Politecnico di Milano, Milano 20133, Italy}
\newcommand{\MIUni}{Physics Department, Universit\`a degli Studi di Milano, Milano 20133, Italy}
\newcommand{\MSU}{Skobeltsyn Institute of Nuclear Physics, Lomonosov Moscow State University, Moscow 119234, Russia}
\newcommand{\NAINFN}{INFN Napoli, Napoli 80126, Italy}
\newcommand{\NAUniDIST}{Department of Structures for Engineering and Architecture, Universit\`a degli Studi ``Federico II'' di Napoli, Napoli 80126, Italy}
\newcommand{\NAUniPHY}{Physics Department, Universit\`a degli Studi ``Federico II'' di Napoli, Napoli 80126, Italy}
\newcommand{\NAUniPHARM}{Pharmacy Department, Universit\`a degli Studi ``Federico II'' di Napoli, Napoli 80131, Italy}
\newcommand{\NAUniCHE}{Chemical, Materials, and Industrial Production Engineering Department, Universit\`a degli Studi ``Federico II'' di Napoli, Napoli 80126, Italy}
\newcommand{\NSU}{Novosibirsk State University, Novosibirsk 630090, Russia}
\newcommand{\OACINAF}{INAF Osservatorio Astronomico di Capodimonte, 80131 Napoli, Italy}
\newcommand{\Petersburg}{Saint Petersburg Nuclear Physics Institute, Gatchina 188350, Russia}
\newcommand{\PGUniCBB}{Chemistry, Biology and Biotechnology Department, Universit\`a degli Studi di Perugia, Perugia 06123, Italy}
\newcommand{\PGINFN}{INFN Perugia, Perugia 06123, Italy}
\newcommand{\PIINFN}{INFN Pisa, Pisa 56127, Italy}
\newcommand{\PIUniPHY}{Physics Department, Universit\`a degli Studi di Pisa, Pisa 56127, Italy}
\newcommand{\PNNL}{Pacific Northwest National Laboratory, Richland, WA 99352, USA}
\newcommand{\Princeton}{Physics Department, Princeton University, Princeton, NJ 08544, USA}
\newcommand{\Queens}{Department of Physics, Engineering Physics and Astronomy, QueenÕs University, Kingston, ON K7L 3N6, Canada}
\newcommand{\RHUL}{Department of Physics, Royal Holloway University of London, Egham TW20 0EX, UK}
\newcommand{\RMTreINFN}{INFN Roma Tre, Roma 00146, Italy}
\newcommand{\RMTreUni}{Mathematics and Physics Department, Universit\`a degli Studi Roma Tre, Roma 00146, Italy}
\newcommand{\RMUnoINFN}{INFN Sezione di Roma, Roma 00185, Italy}
\newcommand{\RMUnoUni}{Physics Department, Sapienza Universit\`a di Roma, Roma 00185, Italy}
\newcommand{\SAINFN}{INFN Salerno, Salerno 84084, Italy}
\newcommand{\SNOLabaddress}{SNOLAB, Lively, ON P3Y 1N2, Canada}
\newcommand{\SNOLAB}{SNOLAB, Lively, ON P3Y 1N2, Canada}
\newcommand{\SSUniCHP}{Chemistry and Pharmacy Department, Universit\`a degli Studi di Sassari, Sassari 07100, Italy}
\newcommand{\STFCppd}{Science \& Technology Facilities Council (STFC), Rutherford Appleton Laboratory, Particle Physics Department, Harwell Oxford, Didcot OX11 0QX, UK}\newcommand{\Sussex}{Physics and Astronomy, University of Sussex, Brighton BN1 9QH, UK}
\newcommand{\Temple}{Physics Department, Temple University, Philadelphia, PA 19122, USA}
\newcommand{\TNFBK}{Fondazione Bruno Kessler, Povo 38123, Italy}
\newcommand{\TNTIFPA}{Trento Institute for Fundamental Physics and Applications, Povo 38123, Italy}
\newcommand{\TNUni}{Physics Department, Universit\`a degli Studi di Trento, Povo 38123, Italy}
\newcommand{\TOINFN}{INFN Torino, Torino 10125, Italy}
\newcommand{\TOPoli}{Department of Electronics and Communications, Politecnico di Torino, Torino 10129, Italy}
\newcommand{\TOUni}{Physics Department, Universit\`a degli Studi di Torino, Torino 10125, Italy}
\newcommand{\TRIUMF}{TRIUMF, 4004 Wesbrook Mall, Vancouver, BC V6T2A3, Canada}
\newcommand{\TUM}{Physik Department, Technische Universit\"at M\"unchen, Munich 80333, Germany}
\newcommand{\UB}{Universiatat de Barcelona, Barcelona E-08028, Catalonia, Spain} 
\newcommand{\UCDavis}{Department of Physics, University of California, Davis, CA 95616, USA}
\newcommand{\UCLA}{Physics and Astronomy Department, University of California, Los Angeles, CA 90095, USA}
\newcommand{\UMass}{Amherst Center for Fundamental Interactions and Physics Department, University of Massachusetts, Amherst, MA 01003, USA}
\newcommand{\UOC}{Department of Chemistry, University of Crete, P.O. Box 2208, 71003 Heraklion, Crete, Greece}
\newcommand{\USP}{Instituto de F\'isica, Universidade de S\~ao Paulo, S\~ao Paulo 05508-090, Brazil}
\newcommand{\VTech}{Virginia Tech, Blacksburg, VA 24061, USA}
\newcommand{\london}{Physics, Kings College London, Strand, London WC2R 2LS, United Kingdom}
\newcommand{\WilliamsCollege}{Williams College, Department of Physics and Astronomy, Williamstown, MA 01267 USA}
\newcommand{\Zaragoza}{Centro de Astropart\'iculas y F\'isica de Altas Energ\'ias, Universidad de Zaragoza, Zaragoza 50009, Spain}
\newcommand{\CPPM}{Centre de Physique des Particules de Marseille, Aix Marseille Univ, CNRS/IN2P3, CPPM, Marseille, France}
\newcommand{\SDakota}{School of Natural Sciences, Black Hills State University, Spearfish, South Dakota 57799, USA}
\newcommand{\STFCInterconnect}{Science \& Technology Facilities Council (STFC), Rutherford Appleton Laboratory, Technology, Harwell Oxford, Didcot OX11 0QX, UK}
\newcommand{\AstroCeNT}{AstroCeNT, Nicolaus Copernicus Astronomical Center of the Polish Academy of Sciences, 00-614 Warsaw, Poland}
\newcommand{\UCRiverside}{Department of Physics and Astronomy, University of California, Riverside, CA 92507, USA}
\newcommand{\UniHAM}{Institute of Experimental Physics, University of Hamburg, Luruper Chaussee 149, 22761, Hamburg, Germany}
\newcommand{\UnivAQ}{Universit\`a degli Studi dell’Aquila, L’Aquila 67100, Italy}
\newcommand{\UniversityofEdinburgh}{School of Physics and Astronomy, University of Edinburgh, Edinburgh EH9 3FD, UK}
\newcommand{\Oxford}{University of Oxford, Oxford OX1 2JD, United Kingdom}
\newcommand{\Warwick}{University of Warwick, Department of Physics, Coventry CV47AL, UK}
\newcommand{\Washington}{Center for Experimental Nuclear Physics and Astrophysics, and Department of Physics, University of Washington, Seattle, WA 98195, USA}
\newcommand{\WUT}{Institute of Radioelectronics and Multimedia Technology, Faculty of Electronics and Information Technology, Warsaw University of Technology, 00-661 Warsaw, Poland}
\newcommand{\UCAS}{University of Chinese Academy of Sciences, Beijing 100049, China}

\author{F.~Acerbi}\affiliation{\TNFBK}
\author{P.~Adhikari}\affiliation{\Carleton}
\author{P.~Agnes}\affiliation{\AQGSSI}\affiliation{\AQLNGS}
\author{I.~Ahmad}\affiliation{\AstroCeNT}
\author{S.~Albergo}\affiliation{\CTUNI}\affiliation{\CTINFN}
\author{I.~F.~Albuquerque}\affiliation{\USP}
\author{T.~Alexander}\affiliation{\PNNL}
\author{A.~K.~Alton}\affiliation{\Augustana}
\author{P.~Amaudruz}\affiliation{\TRIUMF}
\author{M.~Angiolilli}\affiliation{\AQGSSI}\affiliation{\AQLNGS}
% ...
\collaboration{The DarkSide-50 and DarkSide-20k Collaboration}\noaffiliation

\date{\today}

\begin{abstract}
Dark matter detection experiments using liquid argon rely on a precise characterization of the ionization response to nuclear recoils, especially in the keV energy range relevant for light dark matter interactions. In this work, we present a comprehensive analysis that combines new measurements from the ReD setup, part of the DarkSide experimental program, with calibration data from DarkSide-50, as well as results from the ARIS and SCENE experiments. These combined datasets enable improved constraints on atomic screening effects in the modeling of the ionization response of liquid argon to nuclear recoils. The analysis is performed within the Thomas--Imel recombination framework adopted in previous DarkSide studies, and is here further constrained by the inclusion of ReD data, which allow the screening function to be determined from calibration measurements.

By including the updated ionization model into the DarkSide-50 analysis framework, we obtain stronger exclusion limits on low-mass WIMP interactions, setting new world-leading constraints in the 1–3~GeV/c$^2$ WIMP mass range. Finally, we recast the sensitivity projections for the upcoming DarkSide-20k detector, demonstrating a significantly enhanced discovery potential for low-mass dark matter candidates.

%
\end{abstract}

\maketitle



\section{Introduction}

The search for dark matter remains one of the most pressing challenges in astroparticle physics and cosmology. Among the various experimental strategies, liquid argon dual-phase time projection chambers (LAr TPCs) have proven to be highly effective detectors for weakly interacting massive particles (WIMPs) scattering off nuclei.  In dual-phase TPCs, particle interactions in the liquid argon produce both scintillation light and ionization electrons. The electrons are drifted upward by an uniform electric field and extracted into the gas phase, where they generate a secondary scintillation signal via electroluminescence. Both light signals are detected by photosensors located at the top and bottom of the TPC, allowing for precise vertex reconstruction and effective particle identification.

In recent years, the potential of this detection technique has been extended to the low-mass regime (in the GeV/c$^2$ scale) by exploiting the ionization channel alone, as the ability to detect single-electron signals enables sensitivity to keV-scale nuclear recoils. A key ingredient for achieving this sensitivity is the accurate modeling of the ionization yield ($Q_y$) from nuclear recoils, which hinges on a detailed description of the atomic collision processes between the recoiling nucleus and the surrounding argon atoms. As a nucleus propagates through liquid argon, its interactions with neighboring atoms are not governed by bare Coulomb forces but are instead modified by electron screening. The electron clouds surrounding each nucleus alter the effective interaction potential, which is described by screening functions (SFs). These functions determine the energy transferred during atomic collisions and, in turn, directly affect the $Q_y$ observed in the detector.

The present work builds upon this framework, modelling the ion--electron recombination probability with the Thomas--Imel box model~\cite{PhysRevA.36.614}. We provide a unified treatment of all available low-energy LAr $Q_y$ measurements, enabling a meaningful discrimination among screening models and a direct impact on the resulting experimental limits and projections.


\section{The ionization yield model}

In a previous study~\cite{DarkSide:2021bnz}, we attempted to constrain these screening models by combining nuclear recoil data from DarkSide-50, starting at energies of about 0.4~keV, with measurements from ARIS~\cite{Agnes:2018mvl} and SCENE~\cite{Cao:2014gns}, two small-scale liquid argon setups that provided mono-energetic nuclear recoils down to 7~keV.  While further dedicated sub-keV nuclear-recoil calibrations would be valuable, the existing in-situ AmC data already provide direct constraints on the ionization response in the low-energy region relevant for this analysis. However, the experimental sensitivity was insufficient to discriminate between alternative theoretical descriptions, namely the SFs proposed by Ziegler \textit{et al.}~\cite{Ziegler:2010bzy} (hereafter ZBL), Molière~\cite{Moliere:1947zza}, and Lenz-Jensen~\cite{Lenz,Jensen}.   

\begin{figure}[t]
  \centering
  \includegraphics[width=0.95\columnwidth]{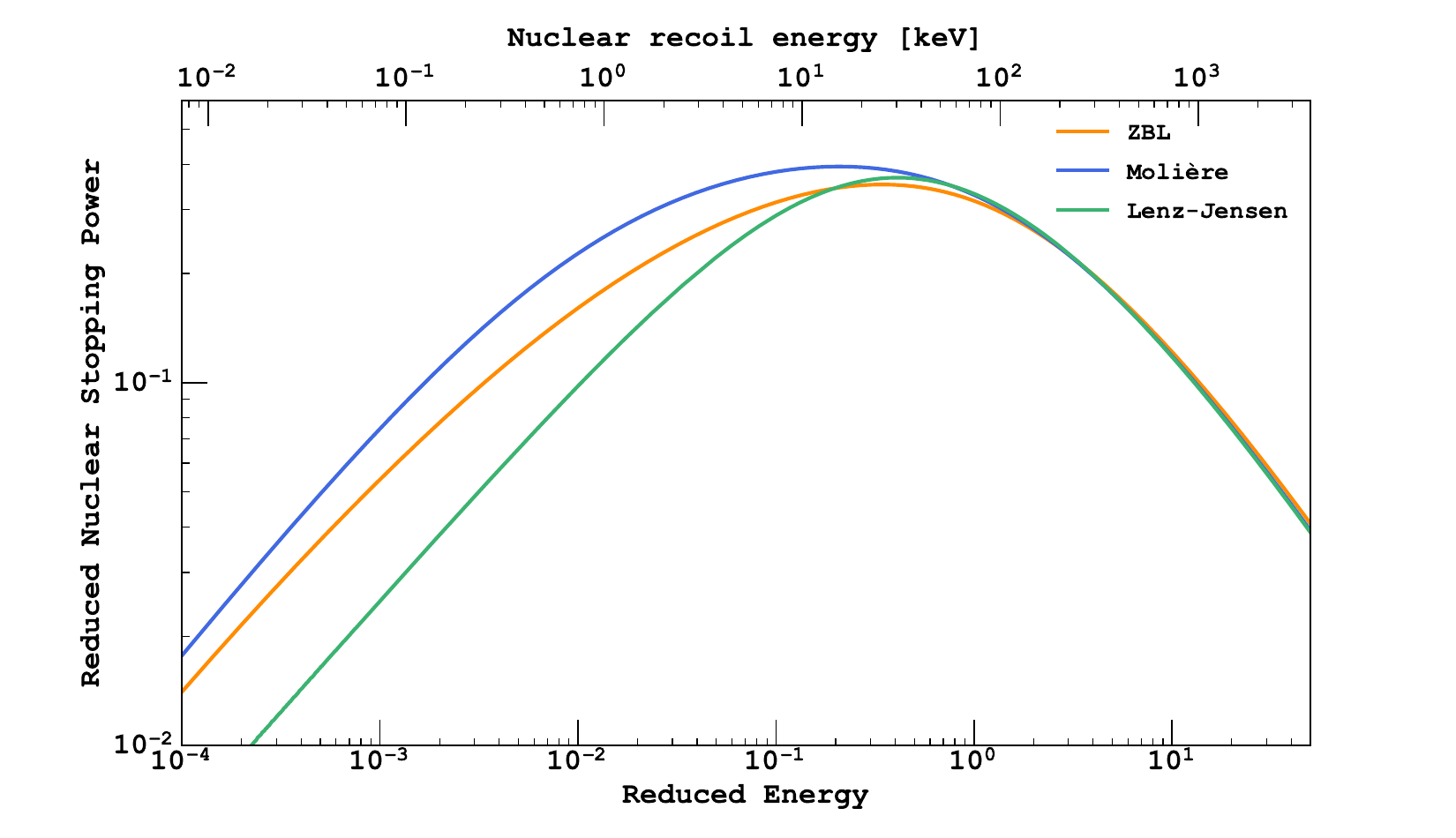}
\caption{
Reduced nuclear stopping power $s_n(\epsilon)$ as a function of the reduced energy $\epsilon$ for the three screening models considered in this work: ZBL, Moli\`ere, and Lenz--Jensen~\cite{Franco:2026tau}. The corresponding nuclear recoil energy scale, shown on the top axis, is computed for liquid argon.
}
  \label{fig:screening}
\end{figure}

These functions, while similar in their general approach to modeling electron cloud effects, differ in their mathematical formulation and underlying assumptions. The ZBL screening function is largely empirical and is widely used in ion-implantation simulations. Both the Molière and Lenz–Jensen screening functions are based on the statistical Thomas–Fermi model. The Molière function provides a three-term exponential approximation to the Thomas–Fermi potential, while the Lenz–Jensen function offers a simpler, single-term analytic form derived from the same theoretical framework. However, it is important to note that in our previous DarkSide-50 analysis~\cite{DarkSide:2021bnz}, as well as in Bezrukov~\textit{et al.}~\cite{Bezrukov:2010qa}, the Molière potential was implemented using the approximate form compiled by Winterbon~\cite{Winterbon1972}, with numerical coefficients incorrectly transcribed in the widely used tabulation by Sigmund~\cite{Sigmund2014}. We also note that, unlike the case of the Lenz–Jensen potential, the Winterbon approximation deviates significantly from the true Molière stopping power~\cite{Winterbon1972}, even when the correct coefficients are used. To address this, we adopt an improved parametrization by Wilson~\textit{et al.}~\cite{Wilson1977}, providing a percent-level agreement with the original Molière screening function.  The three screening models, illustrated in Fig.~\ref{fig:screening}, show noticeable differences at low reduced energies, corresponding to nuclear recoil energies in liquid argon below $\sim$50~keV.

. 

%
%
%
%
%

The limited ability to discriminate between screening models has led  to sizeable differences predictions for $Q_y$ from nuclear recoils, particularly below 5~keV, the  energies most relevant for light WIMP detection. This, in turn, has impacted the interpretation of experimental data and constrained the sensitivity of dark matter searches using liquid argon detectors. In our previous studies, we therefore adopted the ZBL SF, which yields the lowest ionization response among the models considered, namely Lenz–Jensen and Molière as defined in Ref.~\cite{Bezrukov:2010qa}, resulting in the most conservative sensitivity to WIMP interactions for both DarkSide-50~\cite{DarkSide-50:2022qzh}  and DarkSide-20k~\cite{DarkSide-20k:2024yfq} in the GeV/c$^2$ mass range.

The ionization yield model,  $\Qti(E_{nr})$,  describes the number of electrons escaping recombination with ions per unit of recoil energy ($E_{nr}$) as~\cite{DarkSide:2021bnz}
\begin{equation}
\Qti(E_{nr}) = \frac{(1 - r)\, N_i}{E_{nr}},
\label{eq:qy}
\end{equation}

where $N_i$ is the number of initially produced electron–ion pairs and $r$ is the recombination probability. In the low-energy regime (few~keV) we model $r$ with the Thomas–Imel box formalism~\cite{PhysRevA.36.614}, which describes geminate recombination of electrons and ions initially distributed within a small volume ("box") and accounts for the competition between Coulomb attraction, diffusion and electron escape under the applied drift field. In this limit
\begin{equation}
r = 1 - \frac{1}{\gamma N_i}\ln(1+\gamma N_i),
\end{equation}

\noindent where $\gamma = C_{\mathrm{box}}/F$ is the recombination parameter, with $F$ the drift field and $C_{\mathrm{box}}$ a phenomenological parameter that effectively sets the "box" scale.

The initial number of ionization pairs, $N_i$, is estimated assuming a constant excitation-to-ionization ratio, and is modeled as
\begin{equation}
N_i = \beta \, \kappa(\epsilon) = \beta \, \frac{\epsilon \, s_e(\epsilon)}{s_n(\epsilon) + s_e(\epsilon)},
\end{equation}

\noindent where $\beta$ is a normalization constant. The function $\kappa$ quantifies the energy fraction going into electronic excitations, while $s_e$ and $s_n$ are the electronic and nuclear stopping powers, respectively. Both $s_e$ and $s_n$ depend on the dimensionless variable $\epsilon$ ($\sim$0.0135 $E_{nr}/\mathrm{keV}$ in LAr), which in turn depends on the recoil energy $E_{nr}$, the atomic number, and the Thomas-Fermi screening length, as discussed in~\cite{Bezrukov:2010qa}.

It is worth noting that the ionization response model of Eq.~\ref{eq:qy} depends on the nuclear stopping power through $s_n(\epsilon)$, which in turn depends on the choice of atomic screening potential, as well as on two free parameters: the recombination constant $C_{box}$ and the normalization constant $\beta$. 

The model $\Qti(E_{nr})$ from Eq.~\ref{eq:qy} was previously constrained by a simultaneous fit to ARIS, SCENE and DarkSide-50 (Ref.~\cite{DarkSide:2021bnz}). Here we repeat that fit including the new ReD measurement, which provides direct sensitivity in the few-keV region and tightens constraints on the atomic screening models.

\section{The ReD experimental setup} 
 
In the recent campaign of the ReD experiment~\cite{Agnes:2025rxi}, the setup was exposed to neutrons from a $^{252}$Cf fission source. ReD is a small-scale dual-phase LAr TPC specifically designed to study the ionization response to nuclear recoils. The chamber has an active volume of 5$\times$5$\times$6~cm$^3$, with a 7~mm-thick gas pocket above the liquid. An electric field of 200~V/cm, established between two ITO-coated acrylic windows acting as anode and cathode, allows a maximum electron drift time of 54~$\mu$s. Light signals from both scintillation (S1) in the liquid and electroluminescence (S2) in the gas pocket, induced by drifted ionization electrons, are collected by cryogenic silicon photo-multipliers arranged in two 5$\times$5~cm$^2$ tiles placed above and below the active volume. The nuclear-recoil energy estimator S2 is converted into the number of ionization electrons by normalizing it to $g_2 = 18.6\pm0.7$~pe/e$^-$~\cite{redpaper}. Here $g_2$ is defined as the average number of photoelectrons detected per free electron extracted into the gas phase. Effects that reduce the number of electrons surviving drift and extraction in the gas phase are included in the detector response implemented in the Monte Carlo. A more detailed description of the ReD detector is provided in~\cite{Agnes:2021zyq}.

To perform the neutron time-of-flight (ToF) measurements, the setup includes two fast BaF$_2$ scintillators coupled to photo-multiplier tubes, which were placed symmetrically around the $^{252}$Cf source to detect prompt fission radiation and define the start time. The stop signal is provided by a neutron spectrometer located approximately 1~m from the TPC and 2~m from the source. It consists of two 3$\times$3 arrays of 1-inch EJ-276 plastic scintillators, allowing for position reconstruction and neutron/$\gamma$ discrimination via pulse shape analysis.  The arrays are placed symmetrically above and below the beam axis, covering a range of scattering angles of 12° to 17°, which enables exploration of the LAr response to nuclear recoil energies in the [2, 8]~keV range. This is made possible by the broad energy spectrum of neutrons from $^{252}$Cf, which extends to over 10~MeV, with an average energy of approximately 2~MeV.

The overall ToF resolution, measured using prompt $\gamma$s from $^{252}$Cf, is 0.7~ns, allowing the neutron energy to be reconstructed with an uncertainty of 1-2\%. The resulting $Q_y$ is in excellent agreement with that measured by the ARIS experiment~\cite{Agnes:2018mvl} at 7.1~keV, as reported in~\cite{redpaper}, which also provides details of the ReD measurement.

\begin{figure}
  \includegraphics[width=0.99\columnwidth]{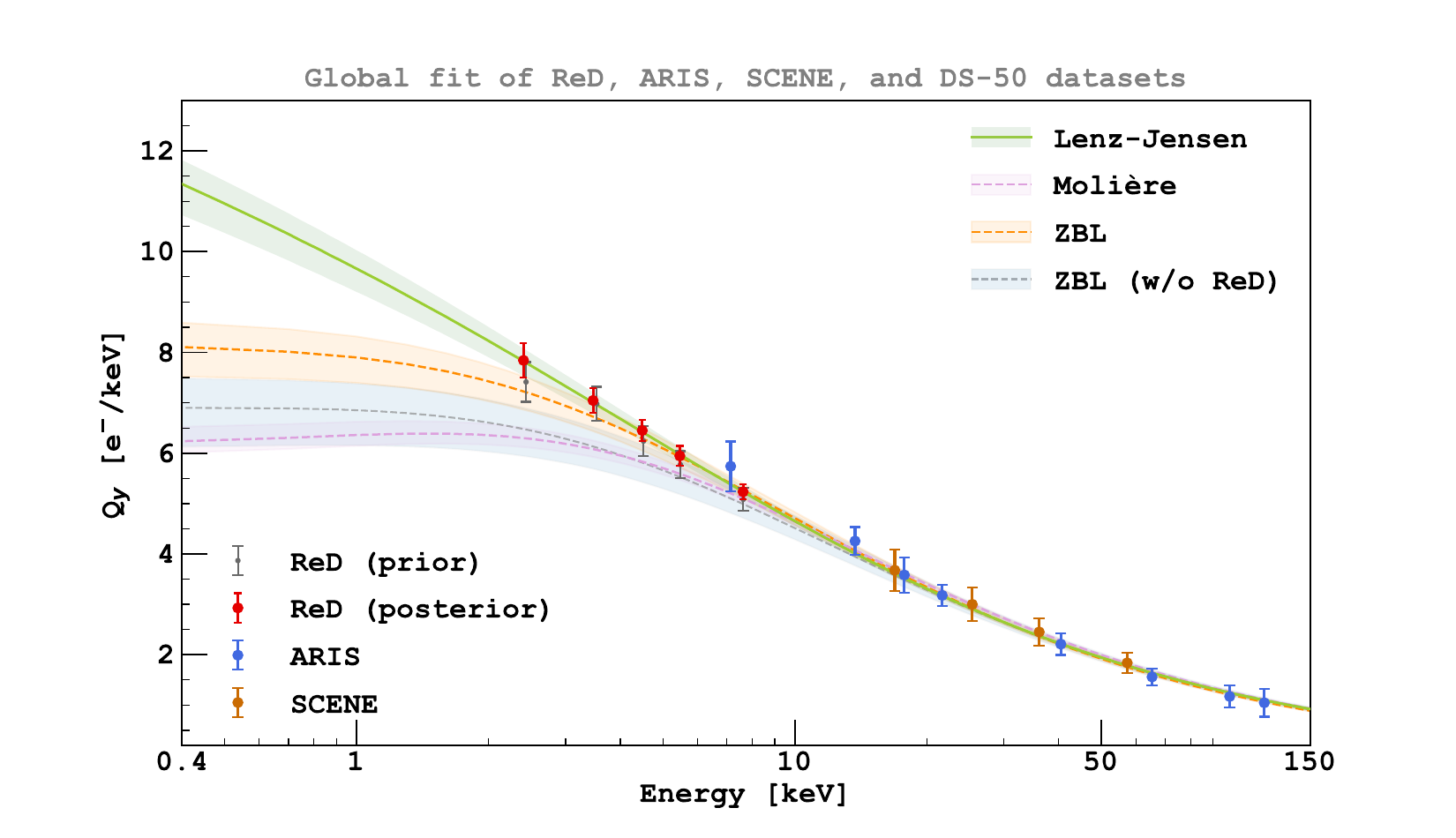}
  \caption{Simultaneous fit to the ReD, ARIS, SCENE, and DarkSide-50 datasets assuming the Lenz-Jensen screening function (green solid line). ReD data points are shown with both prior (gray) and posterior (red) uncertainties. The gray line and its corresponding uncertainty band represent the previous ionization model, based on the ZBL screening function and fitted without ReD data~\cite{DarkSide:2021bnz}. For comparison, the global fit including the ReD dataset was also performed using the screening functions of ZBL (orange dashed line) and Molière (purple dashed line).} 
  \label{fig:qy}
\end{figure}

The $Q_y$ measured by ReD, and shown in Fig.~\ref{fig:qy} (gray points), exhibit an overall uncertainty of 5.7\% at 2.4~keV, decreasing to 4.3\% at 7.5~keV. The dominant sources of systematic uncertainty stem from the calibration of $g_2$ and from a potential vertical shift ($\Delta z$) of the TPC. The latter refers to an offset  between the TPC and the center of the cryostat, which is aligned using a laser with the axis defined by the $^{252}$Cf source and the center of the neutron spectrometer. The presence of non-zero $\Delta z$  was investigated using calibration data from a $^{241}$Am source placed at the vertical center of the cryostat surface. The distribution along the $z$-axis of 59.5~keV $\gamma$-rays from $^{241}$Am, reconstructed in the TPC via the electron drift time, was compared to Monte Carlo simulations. This study yielded an estimated offset of 0.23 $\pm$ 0.96~cm.

The systematic uncertainty on $g_2$ impacts  $Q_y$ uniformly across the energy range, introducing a 3.8\% scaling uncertainty in a consistent direction. In contrast, non-zero $\Delta z$ affects the reconstructed scattering angle and, consequently, the inferred recoil energy of each event, leading to an uncertainty on the mean energy that varies between 0.7\% and 2.4\%. Although this uncertainty is small, a non-zero offset introduces a visible bias in $Q_y$, with opposite trends for nuclear recoils detected by the top and bottom plastic scintillator arrays.

\section{The global fit}


The fit of $\Qti(E_{\text{nr}})$ follows the approach developed in Ref.~\cite{DarkSide:2021bnz}, where a simultaneous fit to data from the ARIS, SCENE, and DarkSide-50 calibrations was performed, providing coverage over the full range from about 3 to more than 200 extracted electrons. The fit minimizes a global $\chi^2$, obtained by summing the $\chi^2$ contributions from each dataset after marginalizing over the corresponding $g_2$ values.

The DarkSide-50 dataset consists of a continuous nuclear recoil spectrum from 0.4 to 200~keV, produced using a neutron source based on the $^{12}\text{C}(\alpha, n)^{15}\text{O}$ reaction, initiated by alpha particles emitted in the decay of $^{241}$Am (AmC source). The SCENE and ARIS datasets provide monoenergetic nuclear recoil lines in the ranges 17--60~keV and 7--120~keV, respectively. In the ARIS detector, which did not include a gas pocket, only the prompt scintillation light (S1) was observed. ARIS data, acquired at 200~V/cm, are rescaled to the DarkSide-50 response by comparing the field-off S1 light yields of the two detectors. This mapping allows each ARIS recoil energy to be associated with the corresponding S1 value at 200~V/cm in DarkSide-50. Further details can be found in Ref.~\cite{DarkSide:2021bnz}.

In this framework, the low-energy ionization response down to sub-keV nuclear recoils is constrained by in-situ DarkSide-50 calibration data, as demonstrated in previous analyses~\cite{DarkSide:2021bnz, DarkSide-50:2022qzh}. In this work, we extend this approach by including the ReD $Q_y$ measurements, which provide direct constraints in the 2--10~keV range, an energy region highly sensitive to the choice of screening potential.

The fit is performed by minimizing a global $\chi^2$ map, numerically constructed as the sum of the individual $\chi^2$ contributions from each dataset. The free parameters $C_{box}$ and $\beta$ are varied, while the drift field is fixed at 200~V/cm, common to all datasets. The ReD nuisance parameters, $g_2$ and the TPC vertical offset, are included in the fit with Gaussian penalty terms based on their respective uncertainties. To account for the systematic uncertainty on $\Delta z$, the ReD data are split into two subsamples corresponding to events tagged by the top and bottom scintillator arrays, which are required to yield consistent $Q_y$ values. Each subsample is independently divided into five energy intervals, and in each interval an unbinned likelihood fit is performed using a Gaussian signal over a constant background. The stability of the extracted $Q_y$ values was verified by varying the energy intervals.

The resulting $2\times5$ $Q_y$ points are fitted with the $\Qti(E_{nr})$ model. For a given screening function model, the fit is performed by scanning the $(C_{box}, \beta)$ parameter space; at each point, the nuisance parameters $g_2$ and $\Delta z$ are marginalized to compute the corresponding $\chi^2$, which contributes to the global $\chi^2$ map.


The posterior distributions are derived from a Bayesian analysis based on the $\chi^2$ maps obtained in the global fit. The likelihood function is constructed as $\mathcal{L} \propto \exp(-\chi^2/2)$ over a uniform prior in the $(C_{\mathrm{box}}, \beta)$ parameter space. Posterior means and confidence regions are estimated via numerical marginalization.

To quantify the preference among the three screening models, we compute the Bayes factor (BF), which compares the marginal likelihoods (or evidences) integrated over the parameter space. This approach is well suited for non-nested models, where standard $\Delta\chi^2$ tests do not apply. Assuming equal prior belief in two models, a $\log_{10} \mathrm{BF} > 2$ implies that the data make one model at least 100 times more likely than the other, a level considered decisive~\cite{jeffreys1961theory, Kass:1995loi}. The results indicate a decisive preference for Lenz-Jensen over ZBL, with $\log_{10} \mathrm{BF} = 3.8$, and an even stronger preference over Molière, with $\log_{10} \mathrm{BF} = 7.2$.

We verify that the posterior mean values are in excellent agreement with the best-fit points (Table~\ref{tab:fit}) obtained from the $\chi^2$ minimization, indicating that the likelihood surfaces are well-behaved and that the fit is not biased by asymmetric parameter degeneracies. This consistency check provides additional confidence in the robustness of the model selection.

\begin{table}[t]
    \centering
    \renewcommand{\arraystretch}{1.5}
    \setlength{\tabcolsep}{10pt}
    \begin{tabular}{l|c c c}
     \hline
     \hline
     \textbf{Parameter} & \textbf{ZBL} & \textbf{Lenz--Jensen} & \textbf{Moli\`ere} \\
     \hline
     $C_{box}$ [V/cm] & $8.1^{+0.1}_{-0.2}$ & $7.9^{+0.2}_{-0.2}$ & $8.6^{+0.3}_{-0.2}$ \\
     $\beta$ [$\times 10^3$] & $7.0^{+0.3}_{-0.2}$ & $6.5^{+0.1}_{-0.3}$ & $8.8^{+0.4}_{-0.5}$ \\
     $\Delta z$ [cm] &  -0.53$_{-0.10}^{+0.19}$ & -0.58$^{+0.05}_{-0.14}$ &  -0.39$_{-0.24}^{+0.05}$  \\
     $g_2$ [pe/$e^-$] & 18.2$_{-0.4}^{+0.4}$ & 18.8$\pm0.4$ & 17.6$_{-0.4}^{+0.4}$\\
     
     \hline
     \hline
    \end{tabular}
    \caption{Best-fit values of the free parameters $C_{\text{box}}$ and $\beta$ from the global fit to all datasets, for each of the tested screening functions.}
    \label{tab:fit}
\end{table}


A simultaneous fit of all datasets was performed separately for each of the three screening functions. The results are shown in Fig.~\ref{fig:qy}, where the ReD $Q_y$ measurements are displayed assuming both the prior and the posterior nuisance parameters, the latter obtained from the fit using the Lenz-Jensen SF. For visualization purposes, the ReD $Q_y$ points shown in the figures correspond to the combined top and bottom array subsamples.

This fit yields nuisance parameters $g_2 = 18.8\pm0.4$~pe/e$^-$ and $\Delta z = -0.58^{+0.05}_{-0.14}$~cm, which are consistent within $1\sigma$ with the priors and with the posteriors obtained using the other two SFs. This demonstrates that the nuisance parameters are stable and largely independent of the specific choice of SF.

With the posterior nuisance parameters, the uncertainty on the ReD $Q_y$ measurement is reduced to 4.5\% at 2.4~keV and 3.0\% at 7.6~keV. This behavior reflects the fact that variations of the nuisance parameters $\Delta z$ and $g_2$ affect the reconstructed recoil energy and electron yield, respectively, and therefore propagate to both the mean value and the uncertainty of the extracted $Q_y$ points.

The global fit using the Lenz-Jensen SF shows excellent agreement with the data. In contrast, both the ZBL-based model~\cite{DarkSide:2021bnz} and the Molière one underestimate $Q_y$ with respect to ReD data below 5~keV.

For the Lenz-Jensen model, the posterior mean and standard deviation are $C_{\mathrm{box}} = 7.9$~V/cm with $\sigma_{C_{\mathrm{box}}} = 0.3$~V/cm and $\beta = 6.6\times10^3$ with $\sigma_{\beta} = 0.3\times10^3$, with a correlation coefficient of 0.24. These values characterize the $Q_y$ model favored by the global fit.

\begin{figure}
  \centering
  \includegraphics[width=0.45\textwidth]{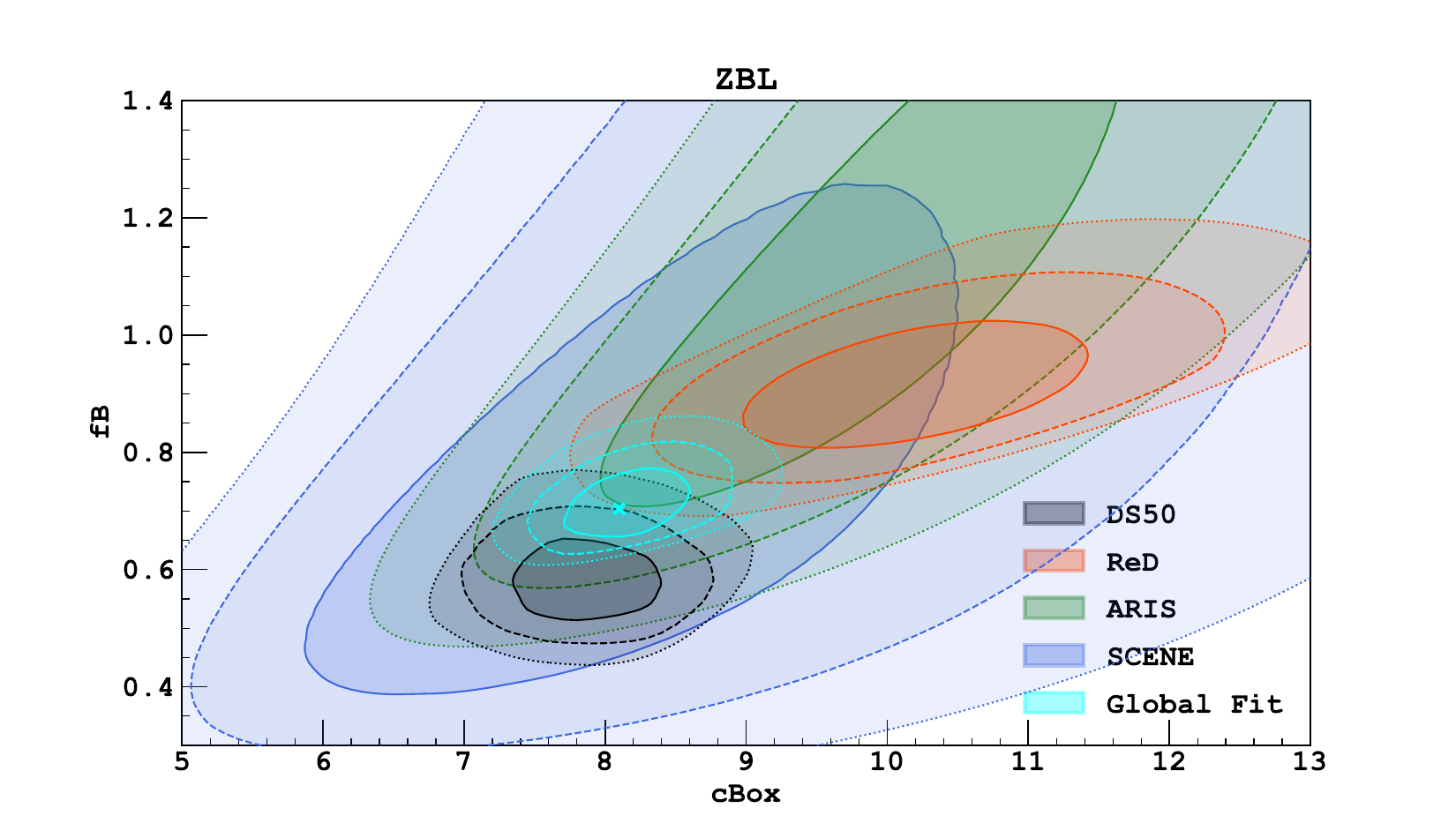}
  \caption{Global fit under the ZBL screening model. The $1\sigma$ (solid line), $2\sigma$ (dashed), and $3\sigma$ (dotted) confidence contours are shown for each dataset, along with the global fit result (cyan).}\label{fig:zbl}
  \includegraphics[width=0.45\textwidth]{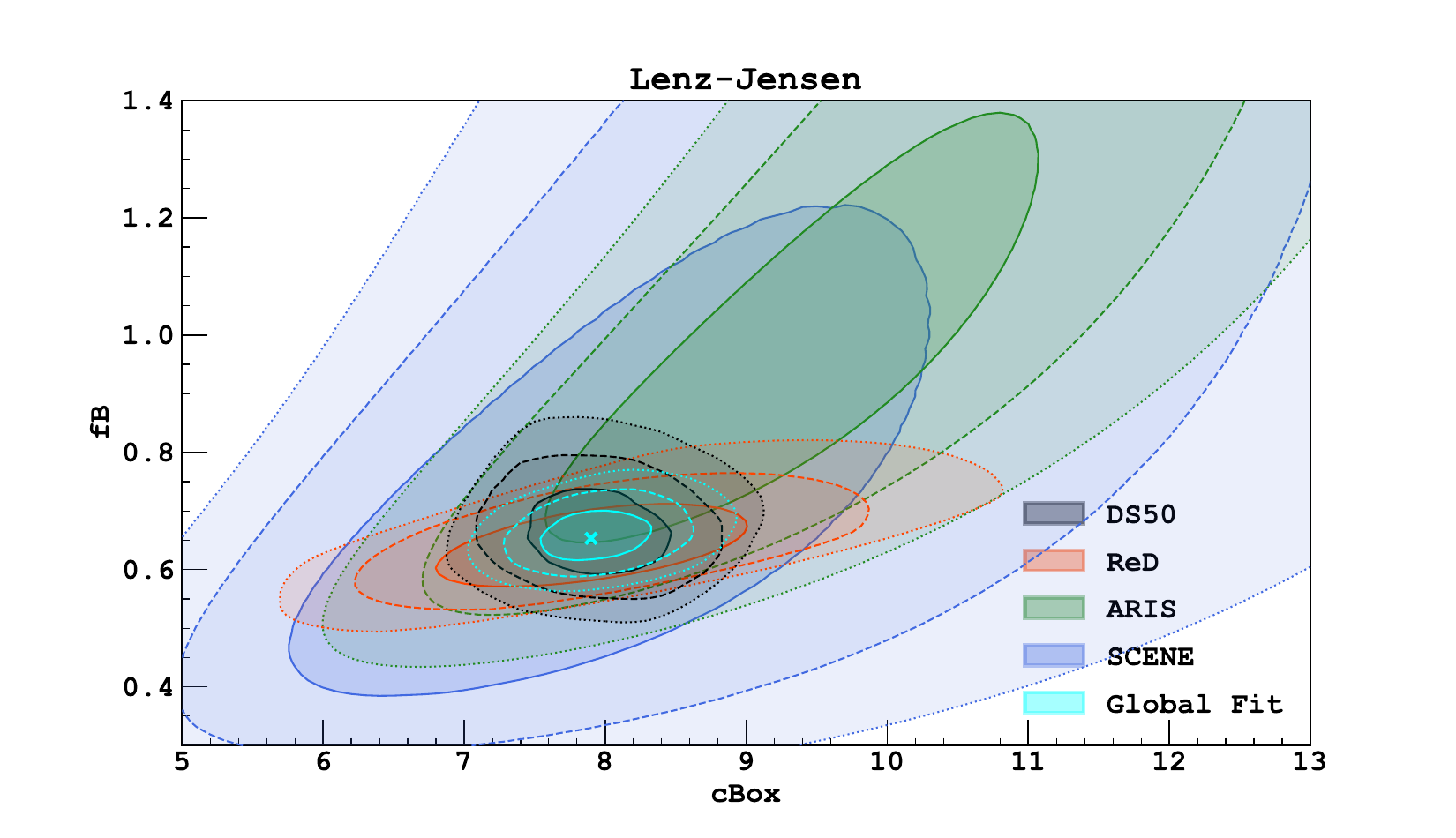}
  \caption{Same as above, for the Lenz-Jensen model. The confidence contours exhibit excellent agreement among all datasets in the $(C_{\mathrm{box}}, \beta)$ plane.}\label{fig:lj}
  \includegraphics[width=0.45\textwidth]{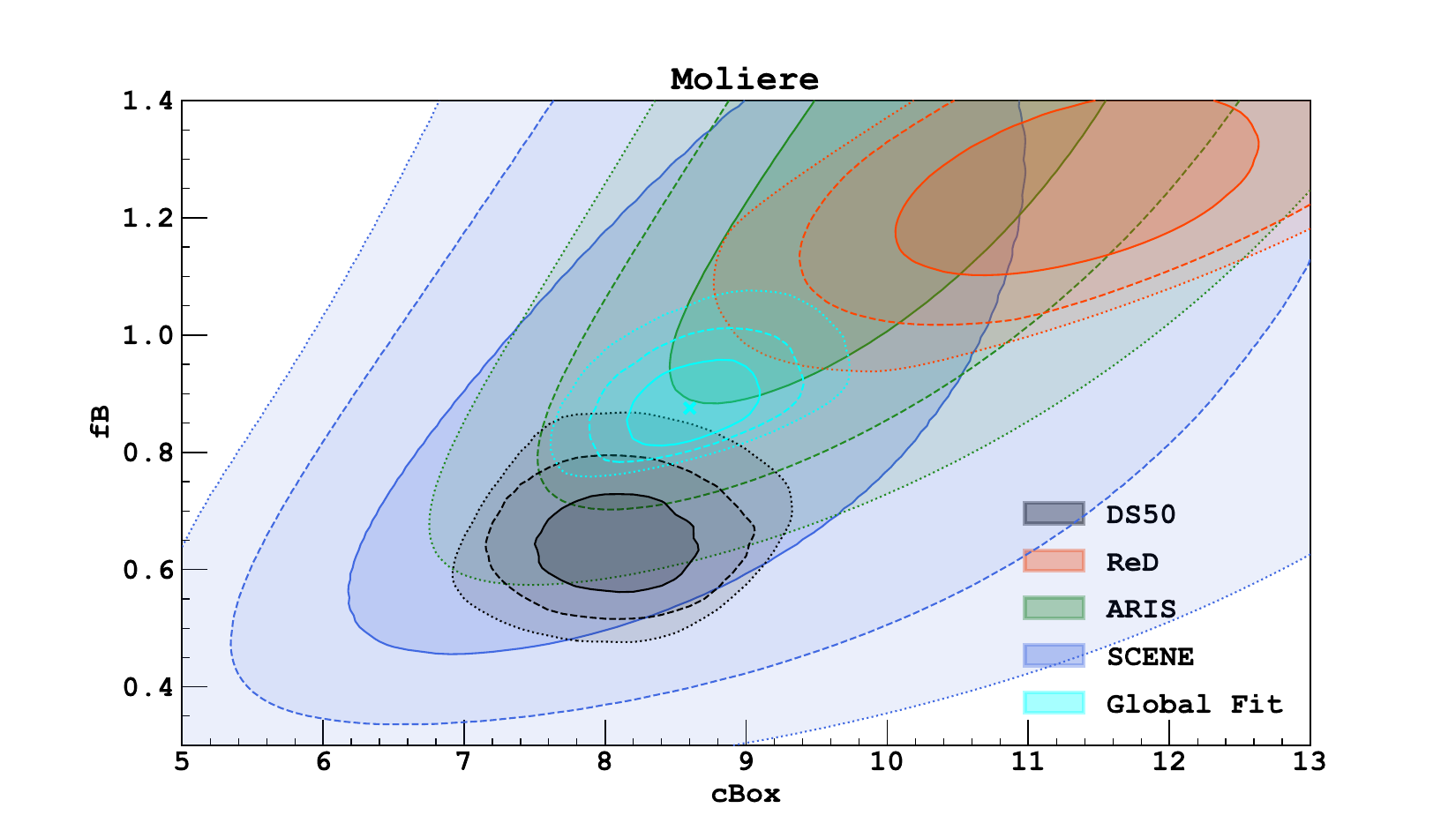}
  \caption{Same as above, for the Molière screening model. Note the strong tension between ReD and DarkSide-50 datasets.}
  \label{fig:allmodels}\label{fig:moliere}
\end{figure}

The contours representing the $1\sigma$, $2\sigma$, and $3\sigma$ confidence levels in the $(C_{\mathrm{box}}, \beta)$ plane are shown in Figs.~\ref{fig:zbl}, \ref{fig:lj}, and \ref{fig:moliere} for the ZBL, Lenz-Jensen, and Molière screening functions, respectively. Each dataset is shown separately, together with the global fit result. The Lenz-Jensen model yields a remarkable agreement across all datasets, with overlapping confidence regions and a significantly higher marginal likelihood compared to the other screening models. In contrast, the ZBL potential exhibits clear tension, while the Molière potential shows even more pronounced inconsistencies, in particular between the DarkSide-50 and ReD datasets.

\begin{figure}
  \includegraphics[width=0.99\columnwidth]{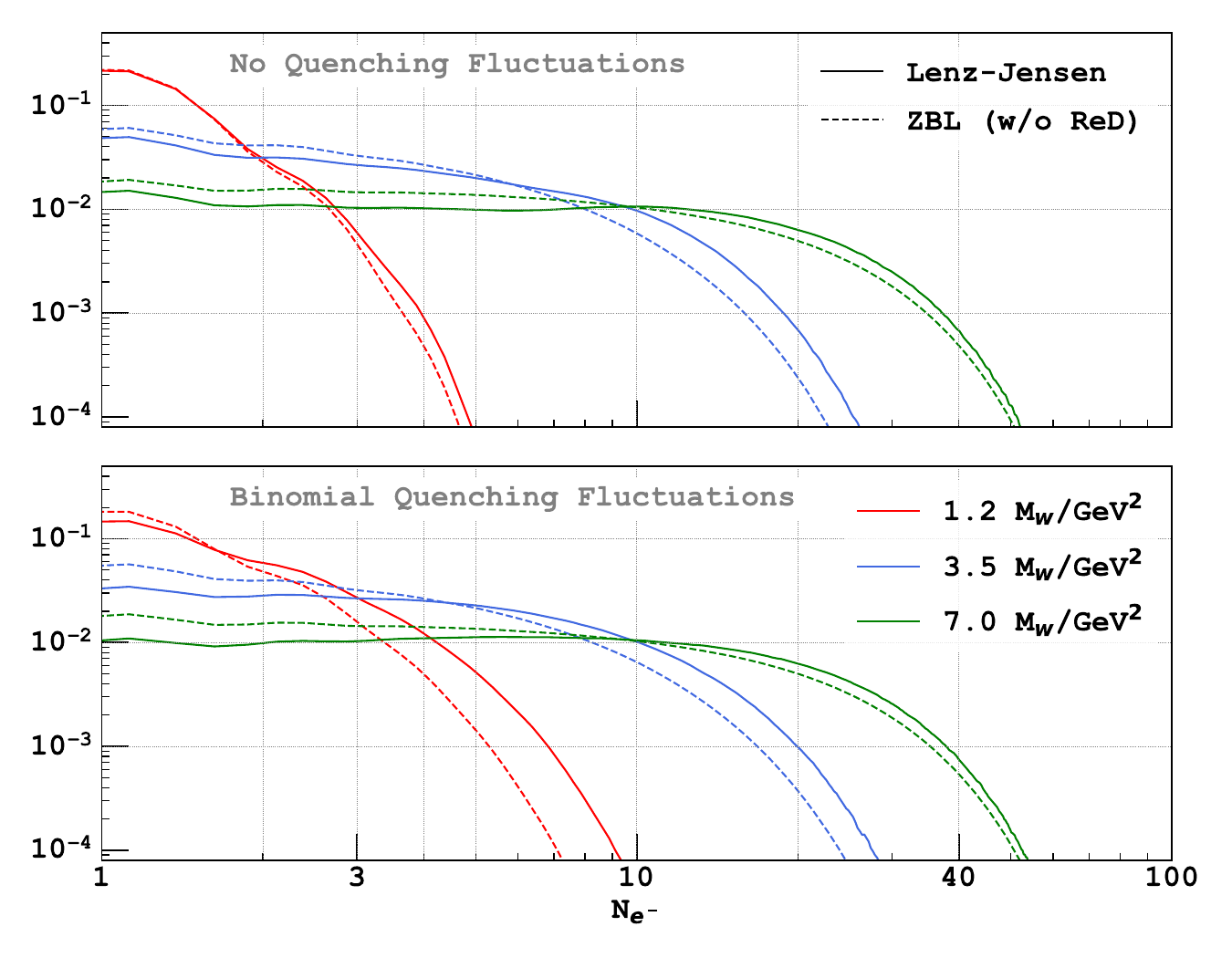}
  \caption{
  Probability density functions of the expected WIMP-induced ionization spectra in DarkSide-50 for WIMP masses of 1.2, 3.5, and 7.0~GeV/c$^2$, shown for three different SF models. The $\Qti(E_{nr})$ response based on the ZBL SF corresponds to the previous fit without ReD data~\cite{DarkSide:2021bnz}, while the Molière SF-based $\Qti(E_{nr})$ curve is from this work. In the top panel, no fluctuations are assumed in the nuclear recoil quenching process (NQ), whereas in the bottom panel, quenching fluctuations are modeled with a binomial distribution (QF).
  } 
  \label{fig:wimps}
\end{figure}

\section{Impact on the WIMP sensitivity}

The impact of the fitted Lenz-Jensen SF-based $\Qti(E_{nr})$, compared to the previous model based on the ZBL SF, on the resulting WIMP spectra is illustrated in Fig.~\ref{fig:wimps}, for various WIMP masses and assuming the DarkSide-50 detector resolution\cite{DarkSide-50:2022qzh}. Due to the absence of a stochastic model for the energy quenching process in nuclear recoils, two scenarios are considered: one assuming no fluctuations (NQ), and one incorporating binomial fluctuations between detectable and undetectable quanta (QF).

The Lenz-Jensen SF predicts a higher ionization yield at low energies compared to the ZBL SF, thereby increasing the probability of detecting signals above analysis thresholds. This results in improved predicted sensitivity, especially for WIMPs with masses in the order of $\mathcal{O}(1~\mathrm{GeV}/c^2)$, which produce nuclear recoils close to the DarkSide experiment thresholds. We then recalculate both the DarkSide-50 observed and the DarkSide-20k expected exclusion limits, adopting the Lenz–Jensen screening function in place of the ZBL one, given its improved agreement with the overall calibration data.

The 90\% C.L. exclusion limits are derived from a binned profile-likelihood fit, following the same procedure and inputs (namely, statistical and systematic uncertainties, background models, and detector resolutions), as in the previous limit calculations for DarkSide-50~\cite{DarkSide-50:2022qzh} and DarkSide-20k~\cite{DarkSide-20k:2024yfq}. For DarkSide-50, the analysis threshold is set at 4~$e^-$, with a dataset corresponding to an effective exposure of approximately 12~ton~day after selection cuts. For DarkSide-20k, the simulated dataset assumes 10 years of data taking, yielding an effective exposure of 342~ton~year and a 2~$e^-$ analysis threshold. The DarkSide-20k background model includes spurious electrons, which dominate the event rate in the few $e^-$ range. Although their origin is not yet fully understood, they are observed to correlate with impurity concentrations in LAr. See refs.~\cite{DarkSide-50:2025umf, DarkSide-50:2022qzh, DarkSide-20k:2024yfq} for additional details.

\begin{figure}
  \includegraphics[width=0.99\columnwidth]{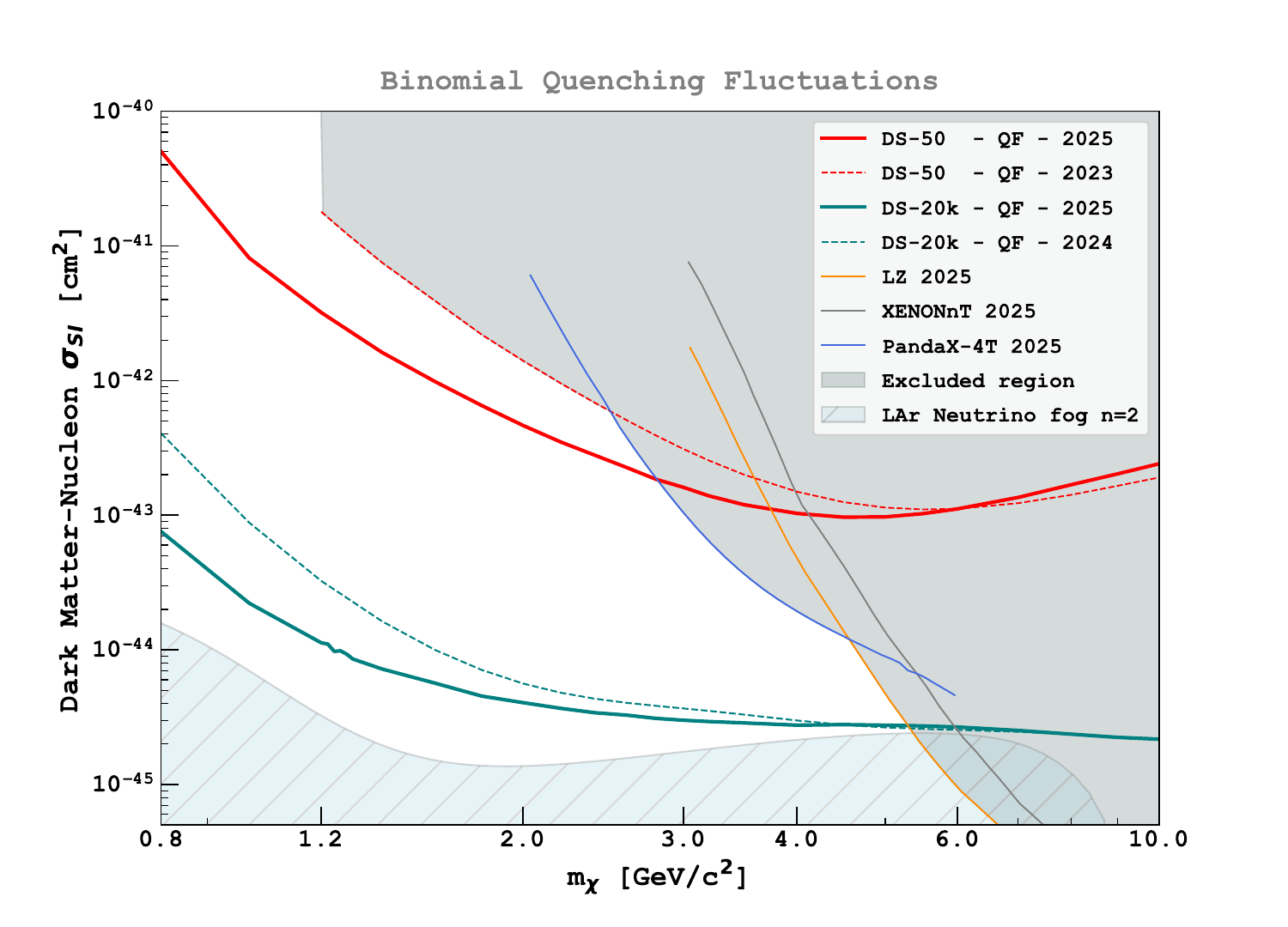}
  \caption{DarkSide-50 (red) exclusion limits with 4 $e^-$ analysis threshold and DarkSide-20k (teal) expected sensitivity assuming binomial quenching fluctuation model (QF),  10 years exposure, and 2 $e^-$  threshold. The updated limits are derived using the Lenz-Jensen screening function in the LAr ionization response model and are compared to previous results obtained with the ZBL screening function. Most recent limits from XENONTnT~\cite{XENON:2024hup}, LZ~\cite{LZ:2025igz}, and PandaX-4T~\cite{PandaX:2025rrz} are also presented. The neutrino fog in LAr with index n=2~\cite{OHare:2021utq} is also shown. } 
  \label{fig:limits_qf}
\end{figure}

\begin{figure}
  \includegraphics[width=0.99\columnwidth]{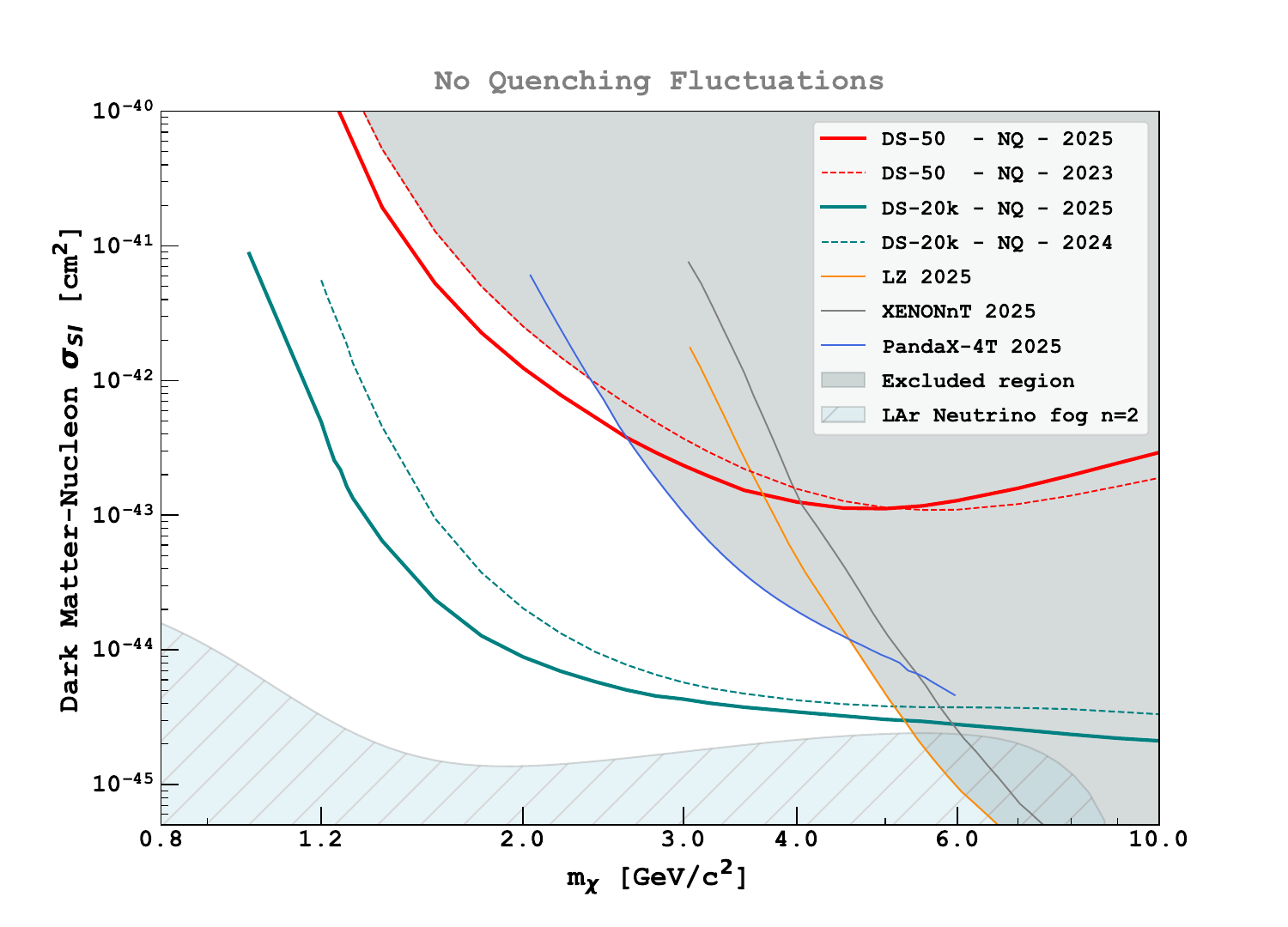}
  \caption{DarkSide-50 exclusion limits (red) and DarkSide-20k projected sensitivity (teal) under the NQ assumption. For color scheme and references, see the caption of Fig.~\ref{fig:limits_nq}. } 
  \label{fig:limits_nq}
\end{figure}

The DarkSide-50 and DarkSide-20k limits are shown in Figs. \ref{fig:limits_qf} and \ref{fig:limits_nq} for the QF and NQ scenarios, respectively. Owing to the improved ionization response model in LAr, the DarkSide-50 limit improves by a factor 5 (2.5) at 1.2~GeV/c$^2$ assuming QF (NQ) fluctuations, setting the world’s most stringent limits in the [0.8, 3.5]~GeV/c$^2$ ([1.0, 3.3]~GeV/c$^2$) mass range. Similarly, for DarkSide-20k, the 90\% C.L. exclusion sensitivity improves by a factor 3 (10) at 1.2~GeV/c$^2$ for the QF (NQ) model.

\section{Conclusions}

The combined ReD, ARIS, SCENE, and DarkSide-50 datasets yield a significantly improved description of the LAr ionization response at low energies, which translates into stronger WIMP exclusion limits and enhanced projected sensitivity for DarkSide-20k. The analysis is performed within the Thomas--Imel recombination framework previously adopted in DarkSide studies. The addition of ReD measurements tightens the constraints on the model parameters and enables a data-driven selection of the preferred atomic screening function. Thanks to its high-accuracy event-by-event energy reconstruction, ReD also has the potential to resolve the ambiguity between the two quenching-fluctuation scenarios. A dedicated campaign within the new ReD+ project, employing a $^{252}$Cf source and a deuterium–deuterium neutron generator, is planned to further pursue this investigation.

%


\vspace{1cm}
\begin{acknowledgments}

This work was supported by the U.S. National Science Foundation (NSF) under Grants No. PHY-0919363, No. PHY-1004072, No. PHY-1004054, No. PHY-1242585, No. PHY-1314483, No. PHY-1314507, No. PHY-1352795, No. PHY-1455351, No. PHY-1606912, No. PHY-1622337, No. PHY-1622415, No. PHY-1812482, No. PHY-1812547, No. PHY-2310091, and No. PHY-2310046, as well as the Major Research Instrumentation Grant No. MRI-1429544 and collaborative NSF grants No. PHY-1211308, No. PHY-1314501, and No. PHY-1455351. Additional support was provided by the U.S. Department of Energy (DOE), Office of Science, under Contracts No. DE-FG02-91ER40671, No. DE-AC02-07CH11359, and No. DE-AC05-76RL01830. We acknowledge the Pacific Northwest National Laboratory (PNNL), operated by Battelle for the DOE under Contract No. DE-AC05-76RL01830, and the Fermi National Accelerator Laboratory (Fermilab), a DOE Office of Science HEP User Facility managed by Fermi Research Alliance, LLC, under Contract No. DE-AC02-07CH11359.

Support in Italy was provided by the Istituto Nazionale di Fisica Nucleare (INFN) through grants from the Ministero dell’Istruzione, dell’Università e della Ricerca (MIUR) — Progetto Premiale 2013 and Commissione Scientifica Nazionale II. This work also received funding from the Ministero dell’Università e della Ricerca (MUR) through the PRIN2020 project (Grant No. PRIN 20208XN9TZ) and the PRIN2022 project (Grant No. 2022JCYC9E, CUP I53D23000690006), issued under the National Recovery and Resilience Plan (NRRP), Mission 4, Component 2, Investment 1.1, funded by the European Union – NextGenerationEU.

This work received support from the French Institut National de Physique Nucléaire et de Physique des Particules (IN2P3) and from IN2P3–COPIN (Grant No. 20-152). Funding was also provided by the Agence Nationale de la Recherche (ANR) under Grants ANR-22-CE31-0021 (X-ArT) and ANR-23-CE31-0015 (FIDAR).

Support was provided by the Natural Sciences and Engineering Research Council of Canada (NSERC), SNOLAB, and the Arthur B. McDonald Canadian Astroparticle Physics Research Institute. The work was also supported by the Deutsche Forschungsgemeinschaft (DFG, German Research Foundation) under Germany’s Excellence Strategy — EXC 2121: Quantum Universe (Grant No. 390833306).

We acknowledge support from the Spanish Ministry of Science and Innovation (MICINN) through Grants PID2022-138357NB-C22 and and PID2022-138357NB-C21, and the Atracción de Talento Grant 2018-T2/TIC-10494; from the Polish National Science Centre (NCN) under Grants No. UMO-2022/47/B/ST2/02015, No. UMO-2023/51/B/ST2/02099, and No. 2021/42/E/ST2/00331; from the Polish Ministry of Science and Higher Education (MNiSW) under Grant No. 6811/IA/SP/2018; and from the Foundation for Polish Science (FNP) through the International Research Agenda Programme AstroCeNT (MAB/2018/7) funded by the European Regional Development Fund. Additional support was provided by the European Union’s Horizon 2020 research and innovation programme under Grant Agreement No. 952480 (DarkWave).

This work was supported by the São Paulo Research Foundation (FAPESP) under Grants No. 2017/26238-4 and No. 2021/11489-7, and by the National Council   for Scientific and Technological Development (CNPq).

We also acknowledge support from the Chinese Academy of Sciences (Grant No. 113111KYSB20210030) and the National Natural Science Foundation of China (Grant No. 12020101004); from the Interdisciplinary Scientific and Educational School of Moscow University “Fundamental and Applied Space Research”; and from the Ministry of Education and Science of the Russian Federation under Project No. FZWG-2020-0032 (2019-1569).

Support from the Science and Technology Facilities Council (STFC), part of UK Research and Innovation, and from The Royal Society (United Kingdom) is also acknowledged.

Finally, we gratefully acknowledge the invaluable technical and logistical support of the staff at the Laboratori Nazionali del Gran Sasso (LNGS), and the assistance of the Fermilab Particle Physics, Scientific, and Core Computing Divisions during the construction and operation of the DarkSide-50 detector.

\end{acknowledgments}

\bibliography{biblio}

\end{document}